\documentclass[%
 reprint,
 amsmath,amssymb,
 aps,
prb,
]{revtex4-1}

\usepackage{graphicx}
\usepackage{dcolumn}
\usepackage{bm}
\usepackage{color}

\newcommand {\braket}[2]{ \langle #1 | #2  \rangle }

\newcommand {\der}[2]{\frac{d #1}{d #2} }
\newcommand {\pder}[2]{\frac{\partial  #1}{\partial #2} }

\DeclareMathOperator*{\SumInt}{%
\mathchoice%
  {\ooalign{$\displaystyle\sum$\cr\hidewidth$\displaystyle\int$\hidewidth\cr}}
  {\ooalign{\raisebox{.14\height}{\scalebox{.7}{$\textstyle\sum$}}\cr\hidewidth$\textstyle\int$\hidewidth\cr}}
  {\ooalign{\raisebox{.2\height}{\scalebox{.6}{$\scriptstyle\sum$}}\cr$\scriptstyle\int$\cr}}
  {\ooalign{\raisebox{.2\height}{\scalebox{.6}{$\scriptstyle\sum$}}\cr$\scriptstyle\int$\cr}}
}
\begin{document}

\title{Atomic cluster expansion of scalar, vectorial and tensorial properties and including magnetism and charge transfer}

\author{Ralf Drautz}
\affiliation{%
 ICAMS, Ruhr-Universit\"at Bochum, Bochum, Germany
}%

\date{\today}

\begin{abstract}
The atomic cluster expansion (Drautz, Phys. Rev. B 99, 014104 (2019)) is extended in two ways, the modelling of vectorial and tensorial atomic properties and the inclusion of atomic degrees of freedom in addition to the positions of the atoms. In particular, atomic species, magnetic moments and charges are attached to the atomic positions and an atomic cluster expansion that includes the different degrees of freedom on equal footing is derived. Expressions for the efficient evaluation of forces and torques are given. Relations to other methods are discussed. 
\end{abstract}

\maketitle


\section{Introduction\label{sec:Intro}}

Simplified models of the interatomic interaction, such as the Ising and the Heisenberg model, were fundamental for the development of solid state physics and materials science. These models provide tremendous insight and understanding with only very few parameters. Today the computational development and design of novel materials requires models that describe interactions for a specific chemistry. A seamless transfer of the classical, generic concepts to quantitative models seems advisable as this allows one to build on decades of research in physics, chemistry and materials science.    

For example, the cluster expansion \cite{Sanchez84} provides a generalization of the Ising model to arbitrary interactions on the lattice. The spin-cluster expansion \cite{Drautz04-5} gives an equivalent generalization of the Heisenberg model, while the atomic cluster expansion (ACE) \cite{Drautz19} may be seen as a numerically feasible extension of empirical interatomic potentials, such as pair potentials, or the Finnis-Sinclair potential \cite{Finnis84} and the embedded atom method \cite{Daw83}, to a general representation of the interatomic interaction. The lowest order approximation of the atomic cluster expansion is based on pair interactions, a systematic and hierarchical expansion enables including arbitrary multi-body order interactions. The expansion is carried out in multi-atom basis functions, which may be chosen to be orthogonal and complete \cite{Dusson20}, such that an expansion of arbitrary accuracy can in principle be obtained. 

The ACE may be related to recent work on body-ordered expansions \cite{Shapeev16,Oord20}, which separates it from machine learning models that are descriptor-based \cite{Behler07,Bartok2010,Manzhos06,Rupp12,Thompson15}. Because of the completeness of the ACE other descriptors and expansions may be written in the form of an ACE, as already discussed for the symmetry functions in neural network potentials \cite{Behler11} and the smooth overlap of atomic positions (SOAP) descriptor\cite{Bartok2013} in Ref.~\onlinecite{Drautz19}, and as I will show explicitly here for the Moments Tensor Potentials (MTP) \cite{Shapeev16} and the Spectral Neighbor Analysis method Potential (SNAP) \cite{Thompson15} here.

I first discuss the general expansion of scalar, vectorial or tensorial atomic properties in the atomic cluster expansion, which also may be viewed to provide a generalization of the tensorial SOAP descriptor \cite{Grisafi18,Grisafi19} to arbitrary order. I then put the ACE in the context of expansions in cartesian tensors, illustrate how these may be expanded in irreducible basis functions of the rotation group and demonstrate that the resulting expressions from the atomic cluster expansion are significantly more sparse.

State of the art models for the combined modelling of atomic magnetic degrees of freedom or atomic charges and atomic positions often additively combine two models, for example, the embedded atom method and the Heisenberg interaction \cite{Ma08,Tranchida18,Dudarev05} or neural network potentials and long-range electrostatic interactions \cite{Artrith11}. In this paper I will show how the atomic cluster expansion may be used to provide a complete expansion for a unified model that combines simultaneously magnetic, charge and atomic positions degrees of freedom. 

The paper is structured as follows. I will first introduce the degrees of freedom that the expansion will capture and then provide a short overview of the atomic cluster expansion. In preparation for the expansion of vectorial and tensorial properties I discuss invariance with respect to translation, rotation and permutation, and relate spherical to cartesian tensors. The atomic cluster expansion will then be specified for scalar, vectorial or tensorial properties of multi-component materials, including non-collinear magnetism and charge transfer, before expressions for forces and torques will be presented.

\section{Degrees of freedom on the atomic scale}

In the following I assume that variables or degrees of freedom may be associated to atoms. The relation between a degree of freedom and an atom may not be unique, or at least not trivial. While the position of an atom is a well defined, classical variable, assigning charge or a magnetic moment to an atom is more difficult. Here I will not discuss how best to evaluate atomic scale variables, but instead will take it for granted that the assignment of atomic variables has been done already.

A property associated to atom $i$, including scalar properties such as energy $E_i$ or charge $q_i$, vectors such as a magnetic moment $\pmb{m}_i$ or a polarizibility tensor depends on the atomic environment $\pmb{\sigma}$ of atom $i$.  I will describe the atomic environment of atom $i$ by variables or degrees of freedom $\sigma_j$ of other atoms $j$ as well as the state of atom $i$, ${\sigma}_i$ as $\pmb{\sigma} = (\sigma_i; \sigma_1, \sigma_2 , \dots , \sigma_N)$. For example, degrees of freedom of atom $j$ that may contribute to a property of atom $i$ are
\begin{equation}
\sigma_j = ( \mu_j, \pmb{r}_{ji}, q_j, \pmb{m}_j, \pmb{T}_j, \dots) \,, \label{eq:sigmaexample}
\end{equation}
with the atomic species $\mu_j$ and where the other contributions illustrate possible dependencies on scalars, {\it e.g.}, the charge $q_j$, vectors $ \pmb{r}_{ji} = \pmb{r}_{j}  - \pmb{r}_{i} $, where $\pmb{r}_{i}$  and $\pmb{r}_{j}$ are the positions of atoms $i$ and $j$, respectively, and $\pmb{m}_j$, or tensors $\pmb{T}_j$. 
A scalar, vectorial or tensorial property $\pmb{G}_i$ of atom $i$ may then be parameterized as a function of $\pmb{\sigma}$, 
\begin{equation}
\pmb{G}_i = \pmb{G}(\pmb{\sigma}) \,. \label{eq:start}
\end{equation}

\section{Atomic cluster expansion \label{sec:ACE}}

The atomic cluster expansion provides a complete descriptor \cite{Drautz19,Dusson20} for the local environment of an atom that I will discuss for multiple degrees of freedom per atom here.
A scalar product between functions $f(\pmb{\sigma})$ and $g(\pmb{\sigma})$ is defined as
\begin{equation}
\braket{f}{g} = \SumInt d\pmb{\sigma} \, f^*(\pmb{\sigma}) g( \pmb{\sigma})\, \omega(\pmb{\sigma})  \,, 
\end{equation}
where discrete degrees of freedom such as atomic species are summed over, continuous degrees of freedom are integrated over a suitable domain and $\omega(\pmb{\sigma})$ is a weight function that may be required by some degrees of freedom. 

Next complete basis functions that only depend on the degrees of freedom associated to one neighboring atom are introduced. These may in general be non-orthogonal
\begin{equation}
\braket{  \phi_{v}^*(\sigma) }{\phi_{u}(\sigma)} = S_{v u} \,. \label{eq:S}
\end{equation}
I assume that the inverse of $\pmb{S}$ exists and use
\begin{equation}
\phi^{v}(\sigma) = \sum_u S^{-1}_{v u} \phi_{u}(\sigma)  \,. \label{eq:up} 
\end{equation}
Then orthogonality and completeness is written as
\begin{align}
\braket{ \phi^{v}(\sigma)}{ \phi_{u}(\sigma)}   &= \delta_{v u} \,, \label{eq:orth1} \\
\sum_v [\phi^{v}(\sigma)]^* \phi_{v}(\sigma')  &=  \delta( \sigma - \sigma') \delta_{\sigma \sigma'} \,, \label{eq:comp1}
\end{align}
where on the right hand side of the second equation the Dirac delta function holds for continuous degrees of freedom and the Kronecker delta for discrete degrees of freedom. For establishing a hierarchical expansion I further choose $\phi_{0} =1$, which may be understood as an atom without properties, {\it i.e.} the vacuum state. 

The atomic cluster expansion is obtained with different degrees of freedom but otherwise in complete analogy to the equations given in Ref.~\onlinecite{Drautz19}. A cluster $\alpha$ with $K$ elements contains  atom $i$ and $K$ further atoms, $\alpha = (i; j_1, j_2, \dots, j_K)$, where atom $i$ is first but otherwise the order of entries in $\alpha$ does not matter and indices are pairwise different $i \neq j_1 \neq j_2 \neq j_K$.  The vector $\nu = (v_0;v_1, v_2, \dots, v_K)$ contains the list of single-atom basis functions in the cluster, and only single-atom basis functions with $v>0$ are considered in $\nu$. A cluster basis function is then given by
\begin{equation}
\Phi_{\alpha \nu} = \phi_{v_0}(\sigma_i) \phi_{v_1}(\sigma_{j_1}) \phi_{v_2}(\sigma_{j_2}) \dots \phi_{v_{K}}(\sigma_{j_K}) \,. 
\end{equation}
The orthogonality and completeness of the single-atom basis functions transfers to the cluster basis functions
\begin{align}
\braket{\Phi^{\alpha \nu}}{\Phi_{\beta \mu}} = \delta_{\alpha \beta} \delta_{\nu \mu} \,, \label{eq:orth2} \\
1 + \sum_{\gamma \subseteq \alpha} \sum_\nu [\Phi^{\gamma \nu}](\pmb{\sigma})^* \Phi_{\gamma \nu}(\pmb{\sigma}')  &=  \delta( \pmb{\sigma} - \pmb{\sigma}') \delta_{\pmb{\sigma} \pmb{\sigma}'}\,, \label{eq:comp2}
\end{align} 
where $\alpha$ is an arbitrary cluster and the right hand side of the completeness relation is the product of the relevant right hand sides of Eq.(\ref{eq:comp1}). The expansion of an element of $\pmb{G}$, Eq.(\ref{eq:start}) may therefore be written in the form
\begin{align}
{G}(\pmb{\sigma}) = J_0 + \sum_{\alpha \nu} J^{\alpha \nu} \Phi_{\alpha \nu}(\pmb{\sigma}) \,,\label{eq:CE}
\end{align}
and the expansion coefficients  $J^{\alpha \nu} $ obtained by projection
\begin{equation}
J^{\alpha \nu} = \braket{\Phi^{\alpha \nu}} {{G}(\pmb{\sigma})}\,.\label{eq:CEJ}
\end{equation}

For convenience and readability I will write the indices of the expansion coefficients as subscripts in the following, with the understanding that all indices of expansion coefficients should be written as superscripts. Writing the expansion Eq.(\ref{eq:CE}) explicitly in single-atom basis functions leads to
\begin{align}
&{G}(\pmb{\sigma}) = \sum_{v_0} J^{(0)}_{v_0}  \phi_{v_0}(\sigma_i) \nonumber \\
&+ \sum_{j}^{i \neq j} \sum_{v_0 v_1} J^{(1)}_{v_0 v_1}  \phi_{v_0}(\sigma_{i}) \phi_{v_1}(\sigma_{j}) \nonumber \\
&+ \frac{1}{2} \sum_{j_1 j_2}^{i \neq j_1 \neq j_2} \sum_{v_0 v_1 v_2} J^{(2)}_{v_0 v_1 v_2 }  \phi_{v_0}(\sigma_{i}) \phi_{v_1}(\sigma_{j_1}) \phi_{v_2}(\sigma_{j_2}) \nonumber \\
&+ \frac{1}{3!} \sum_{j_1 j_2 j_3 }^{i \neq j_1 \neq j_2, \dots} \sum_{v_0 v_1 v_2 v_3}   J^{(3)}_{v_0  v_1 v_2 v_3}   \phi_{v_0}(\sigma_{i}) \phi_{v_1}(\sigma_{j_1}) \phi_{v_2}(\sigma_{j_2}) \phi_{v_3}(\sigma_{j_3}) \nonumber \\
&+ \dots \,. \label{eq:mbint}
\end{align}
This may be rewritten in a slightly different way with unrestricted sums and updated expansion coefficients
\begin{align}
&{G}(\pmb{\sigma}) = \sum_{v} c^{(0)}_{v}  \phi_{v}(\sigma_i) \nonumber \\
&+ \sum_{j} \sum_{v_0 v_1} c^{(1)}_{v_0 v_1}  \phi_{v_0}(\sigma_{i}) \phi_{v}(\sigma_{j}) \nonumber \\
&+ \frac{1}{2} \sum_{j_1 j_2} \sum_{v_0 v_1 v_2} c^{(2)}_{v_0 v_1 v_2 }  \phi_{v_0}(\sigma_{i}) \phi_{v_1}(\sigma_{j_1}) \phi_{v_2}(\sigma_{j_2}) \nonumber \\
&+ \frac{1}{3!} \sum_{j_1 j_2 j_3 } \sum_{v_0 v_1 v_2 v_3}   c^{(3)}_{v_0  v_1 v_2 v_3}   \phi_{v_0}(\sigma_{i}) \phi_{v_1}(\sigma_{j_1}) \phi_{v_2}(\sigma_{j_2}) \phi_{v_3}(\sigma_{j_3}) \nonumber \\
&+ \dots \,, \label{eq:mbint2}
\end{align}
where only $i$ is excluded from the summations over $j_1, j_2, \dots$. As had already been pointed out in  Ref.~\onlinecite{Drautz19}, the expansion Eq.(\ref{eq:mbint2}) is identical to Eq.(\ref{eq:mbint}), with expansion coefficients  $c^{(K)}_{\nu}$ that are different from the expansion coefficients $J^{(K)}_{\nu}$ in Eq.(\ref{eq:mbint}). The expansion coefficients  $c^{(K)}_{\nu}$ are simple functions of $J^{(K)}_{\nu}$ that may be obtained by taking into account that products of basis functions of the same argument may be expanded into linear combinations of single basis functions, for example, $\sum_v a_v \phi_{v}(\sigma_{j}) = \phi_{v_1}(\sigma_{j})\phi_{v_2}(\sigma_{j})$, {\it etc.}, such that the self-interactions are removed by an appropriate modification of a lower-order expansion coefficient. The detailed relation between $J^{(K)}_{\nu}$ and $c^{(K)}_{\nu}$ is given in Ref.~\onlinecite{Dusson20}.

I next introduce the combined atomic density 
\begin{equation}
\rho_i({\sigma})  = \sum_j^{j \neq i} \delta({\sigma} - {\sigma}_j) \,,
\end{equation}
where for the example of Eq.(\ref{eq:sigmaexample}) this implies
\begin{equation}
\delta({\sigma} - {\sigma}_j) = \delta_{\mu \mu_j} \delta( \pmb{r} - \pmb{r}_{ji}) \delta(q - q_j) \delta( \pmb{m} - \pmb{m}_j) \delta( \pmb{T} -\pmb{T}_j) \dots \,,
\end{equation}
and the atomic base is obtained as
\begin{equation}
A_{iv} = \braket{\rho_i}{\phi_{v}}  = \sum_j^{j \neq i} \phi_{v}(\sigma_{j}) \,. \label{eq:AB}
\end{equation}
To capture the properties inherently associated to atom $i$ I further introduce
\begin{equation}
\rho^{(0)}_i({\sigma})  = \delta({\sigma} - {\sigma}_i) \,,
\end{equation}
and
\begin{equation}
A^{(0)}_{iv} =  \braket{\rho^{(0)}_i}{\phi_{v}} =  \phi_{v}(\sigma_{i}) \,.
\end{equation}

Except for $v_0$ it is sufficient to sum over ordered sets of basis functions $v_1 \geq v_2 \geq v_3 \dots$, such that the expansion Eq.(\ref{eq:mbint2}) is written as 
\begin{align}
{G}(\pmb{\sigma}) &=  \sum_{v_0} c^{(0)}_{v_0}  A^{(0)}_{iv_0} + \sum_{v_0 v_1} c^{(1)}_{v_0 v_1}  A^{(0)}_{iv_0} A_{iv_1}  \nonumber \\ 
&+ \sum_{v_0 v_1 v_2}^{v_1 \geq v_2} c^{(2)}_{v_0 v_1 v_2}   A^{(0)}_{iv_0} A_{iv_1} A_{iv_2} \nonumber \\ 
&+  \sum_{v_0 v_1 v_2 v_3}^{v_1 \geq v_2 \geq v_3}   c^{(3)}_{v_0 v_1 v_2 v_3 }   A^{(0)}_{iv_0} A_{i v_1}  A_{i v_2}  A_{i v_3} + \dots \,. \label{eq:mbfinal}
\end{align}
In this way the linear expansion in cluster correlation functions Eq.(\ref{eq:CE}) is expressed as a polynomial in the atomic base $A_{iv}$.

\section{Translation,  rotation,  inversion and permutation \label{sec:trip}}

The degrees of freedom are constructed such that 
the expansion Eq.(\ref{eq:mbfinal}) is immediately invariant under translation.

For many systems we further expect a well-defined transformation under rotation. For example, without applied field we expect that a scalar property such as the energy is invariant under rotation, or in other words, it belongs to the irreducible representation $\mathcal{D}^{(l=0)}$ of the rotation group. A vector, for example, the magnetic moment on atom $i$ transforms according to the irreducible representation $l=1$ of the rotation group, $\mathcal{D}^{(l=1)}$. A tensor of rank two transforms as $\mathcal{D}^{(1)} \times \mathcal{D}^{(1)} = \mathcal{D}^{(2)}  + \mathcal{D}^{(1)}  + \mathcal{D}^{(0)} $, {\it i.e.}, a symmetric matrix with 5 independent matrix elements, an anti-symmetric matrix with 3 independent matrix elements and one constant, the trace, required to characterize the transformation behavior of the 9 matrix elements under rotation. A tensor of rank three transforms as $\mathcal{D}^{(1)} \times \mathcal{D}^{(1)} \times \mathcal{D}^{(1)} = \mathcal{D}^{(3)}  + 2\mathcal{D}^{(2)}  + 3\mathcal{D}^{(1)} + \mathcal{D}^{(0)}$, and higher order tensors accordingly \cite{Brink68}. The relation between cartesian and spherical tensors and their representation in irreducible representations of the rotation group will further be discussed in Sec.~\ref{sec:cartesian}. 

Next I classify the property that is expanded according to its irreducible representation and therefore also characterize the basis functions $\phi_{v}$ according to their properties under rotation, {\it i.e.}, by their irreducible representation $l$.
An irreducible representation $l$ of the rotation group comprises $2l+1$ basis functions, labeled by $m = {-l}, {-l+1}, \dots, {l-1}, l$. A product of two irreducible representations $l_1$ and $l_2$ may be decomposed into irreducible representations $\mathcal{D}^{(L)}$, where exactly one representation with $l_1 + l_2 \geq L \geq |l_1 - l_2|$ is contained in the product. I denote the coupling by $( l_1 l_2)L$. With Clebsch-Gordan coefficients $C^{M m_1 m_2}_{L l_1 l_2}$, the matrix elements are given as
\begin{equation}
\{ (l_1 l_2)L \}_{M m_1 m_2} = C^{M m_1 m_2}_{L l_1 l_2} \,.
\end{equation}
If more than two basis functions are coupled, the extraction of the irreducible representations from the product representation may proceed along different sequences, which implies different intermediate couplings. The generalized Clebsch-Gordan coefficients or the generalized Wigner symbols from products of the Wigner 3j symbol may be used to characterize the coupling sequence \cite{Yutsis62}.

For the coupling of the basis functions their order as well as the coupling scheme is relevant, see Refs.~\onlinecite{Yutsis62},~\onlinecite{SingerFaehnle2006},~\onlinecite{Dusson20} for a detailed discussion. Different orders and different coupling schemes may lead to different product basis functions that are related by unitary transformations and for our purposes equivalent, {\it c.f.} recoupling with the Wigner 6j symbols or the Racah W coefficients for three spins and the Wigner 9j symbols or Fano X coefficients for four spins \cite{Wigner31,Racah42,Fano59,Brink68,Varshalovich88}. I choose a particular coupling of the angular momenta $\pmb{l} = (l_1, l_2, \dots, l_N)$ that consists of iterative pairwise couplings,
\begin{align}
&\left(
\begin{array}{c}
\pmb{l} \\
\pmb{L} 
\end{array}
L_{1 \dots N}
\right)_N 
=
\left( 
\begin{array}{c}
\begin{array}{ccccc}
l_1 & l_2 & l_3 & \dots & l_N
\end{array}\\
\begin{array}{ccc}
L_{12} & L_{34} & \dots
\end{array}
\end{array}
L_{1 \dots N}
\right)
 \nonumber \\
&=  \left(  \left(  \left(l_1 l_2 \right) {L_{12}}  \left( l_3 l_4 \right) {L_{34}} \right) {L_{1234}} \left(l_5 l_6\right) {L_{56}} \dots l_N\right) {L_{123456 \dots N}} \,,
\end{align}
with the intermediate angular momenta $\pmb{L} = (L_{12}, L_{34}, \dots )$ and the resulting angular momentum $L_{123456 \dots N}$. 

Using Clebsch-Gordan coefficients the brackets are written as
\begin{equation}
\left(  \left(l_1 l_2 \right) {L_{12}}  \left( l_3 l_4 \right) {L_{34}} \right) {L_{1234}} = C_{L_{1234} L_{12} L_{34}} C_{L_{12} l_1 l_2}C_{L_{34} l_3 l_4} \,,
\end{equation}
with summation over the intermediate $M_{12},M_{34}$ implied and $m_1, m_2, m_3, m_4, M_{1234}$ suppressed.

For coupling two angular momenta, one has
\begin{align}
\left(
\begin{array}{cc}
\pmb{l} & L_{12} 
\end{array}
\right)_2 
=&\left( 
\begin{array}{ccc}
l_1 &  l_2 & L_{12}  
\end{array}
\right)_2 =    \left(l_1 l_2 \right){L_{12}}  \,.
\end{align}
For coupling three angular momenta,
\begin{align}
\left(
\begin{array}{c}
\pmb{l} \\
\pmb{L} 
\end{array}
L_{123}
\right)_3 
=&\left( 
\begin{array}{c}
\begin{array}{ccc}
l_1 &  l_2 & l_3\\
\end{array}\\
\begin{array}{c}
L_{12} 
\end{array}
\end{array}
L_{123}
\right)_3 =   \left( \left(l_1 l_2 \right){L_{12}} l_3\right){L_{123}}  \,.
\end{align}
For four angular momenta,
\begin{align}
&\left(
\begin{array}{c}
\pmb{l} \\
\pmb{L} 
\end{array}
L_{1234}
\right)_4 
=\left( 
\begin{array}{c}
\begin{array}{cccc}
l_1 &  l_2 & l_3 & l_4\\
\end{array}\\
\begin{array}{cc}
L_{12}  & L_{34}
\end{array}
\end{array}
L_{1234}
\right)_4 \nonumber \\
&=   \left( \left(l_1 l_2 \right){L_{12}}    \left(l_3 l_4 \right){L_{34}}\right){L_{1234}}  \,.
\end{align}
For five angular momenta,
\begin{align}
&\left(
\begin{array}{c}
\pmb{l} \\
\pmb{L} 
\end{array}
L_{12345}
\right)_5 
=
\left( 
\begin{array}{c}
\begin{array}{ccccc}
l_1 &  l_2 & l_3 & l_4 & l_5\\
\end{array}\\
\begin{array}{ccc}
L_{12}  & L_{34} & L_{1234} 
\end{array}
\end{array}
L_{12345}
\right)_5  \nonumber \\
&=  \left( \left( \left(l_1 l_2 \right){L_{12}}    \left(l_3 l_4 \right){L_{34}}\right){L_{1234}}  l_5 \right){L_{12345}} \,.
\end{align}
For six angular momenta,
\begin{align}
&\left(
\begin{array}{c}
\pmb{l} \\
\pmb{L} 
\end{array}
L_{123456}
\right)_6 
=
\left( 
\begin{array}{c}
\begin{array}{cccccc}
l_1 &  l_2 & l_3 & l_4 & l_5 & l_6\\
\end{array}\\
\begin{array}{cccc}
L_{12}  & L_{34} & L_{1234} & L_{56} 
\end{array}
\end{array}
L_{123456}
\right)_6  \nonumber \\
&=  \left( \left( \left(l_1 l_2 \right){L_{12}}    \left(l_3 l_4 \right){L_{34}}\right){L_{1234}}  \left(l_5 l_6 \right){L_{56}}  \right){L_{123456}} \,.
\end{align}

For a rotationally invariant scalar, such as an interatomic potential, on requires $L_{12 \dots N}= 0$. For a vector-valued quantity one requires $L_{12 \dots N} = 1$ and for the  symmetric $\mathcal{D}^{(2)}$ contribution to a rank two tensor $L_{12 \dots N} = 2$, {\it etc}.

Many of the couplings are zero, for example, $(l_1 l_2)0 = 0$ for $l_1 \neq l_2$, $((l_1 l_2){L_{12}} l_3){0} = 0$ for $L_{12} \neq l_3$, $L_{12} < | l_1 - l_2|$ or $ L_{12} > l_1 + l_2$, {\it etc}. The parity of the product representation is given by $(-1)^{(l_1 + l_2 + \dots + l_N)}$, therefore invariance with respect to inversion requires that $l_1 + l_2 + \dots + l_N$ is an even number. This also limits intermediate couplings, for example, two identical basis functions cannot couple to negative parity, $(l l)L = 0$ for $L = 1, 3, 5, \dots$. If two or more of the $l_i$ are identical, the number of intermediate couplings $L$ that result in linearly independent functions is further reduced\cite{Dusson20}. For the important case $L_{12 \dots N}= 0$ the condition $(l_1 l_2)0 = 0$ for $l_1 \neq l_2$ imposes further constraints on possible intermediate couplings.

The possible combinations of $m_1, m_2, \dots m_N$ and intermediate couplings $M_{12}, M_{34}, \dots, M_{12 \dots N}$ are also limited. For example, the Clebsch-Gordan coefficients $\{ (l_1 l_2)L \}_{M m_1 m_2}$ are non-zero only if $ M = m_1 + m_2$. Rotational invariance further requires $m_1 +m_2 + \dots + m_N = 0$.

By construction the atomic cluster is expansion is invariant with respect to permutation of identical atoms.

\subsection{Relation to cartesian tensors and expansions in hyperspherical harmonics \label{sec:cartesian} }

The relation between cartesian and spherical tensors is well established \cite{Brink68,Stone76,Stone84,StoneBook}. Here I give explicit expressions for the expansion of tensor products of unit length vectors $\hat{\pmb{r}}$ in spherical harmonics. I consider tensor products $\hat{\pmb{r}} \otimes \hat{\pmb{r}}$,   $\hat{\pmb{r}} \otimes \hat{\pmb{r}} \otimes \hat{\pmb{r}}$, $\hat{\pmb{r}} \otimes \hat{\pmb{r}} \otimes \hat{\pmb{r}}\otimes \hat{\pmb{r}}$, $\dots$ with matrix elements $\hat{r}_{n_1} \hat{r}_{n_2}$, $\hat{r}_{n_1} \hat{r}_{n_2} \hat{r}_{n_3}$, $\hat{r}_{n_1} \hat{r}_{n_2} \hat{r}_{n_3}\hat{r}_{n_4}$, $\dots$ respectively.

It is then straighfoward to show that a tensor of order $N$ may be represented as a linear combination of spherical harmonics up to angular momentum $N$,
\begin{align}
\hat{r}_{n_1} \hat{r}_{n_2} \dots \hat{r}_{n_N} &= \sum_{L = 0}^N \sum_{M = -L}^L X_{n_1 n_2 n_3 \dots n_N}^{L M} Y_{L}^{M} \,.  \label{eq:Ytor}
\end{align}
The expression for the transformation matrix $\pmb{X}$ is derived in App.~\ref{app:MT} and given by Eq.(\ref{eq:X}). 

The expansion in spherical harmonics provides a sparse representation: the number of matrix elements of the cartesian tensor of order $N$ is given by $3^N$ in three dimensions, while the number of spherical harmonics is given by $1 + 3 + 5 \dots 2N+1 = (N+1)^2$. For example, a tensor of order 10 with $3^{10} = 59049$ matrix elements may be represented by only $121$ spherical harmonics.

One may therefore assume that the evaluation of the spherical harmonics is considerably faster than the evaluation of the tensor products of the same order. Traditionally the spherical harmonics are evaluated in spherical coordinates and the transformation to a spherical coordinate system may be viewed as an overhead for computing spherical harmonics. In  App.~\ref{app:SH} I discuss the computation of spherical harmonics as polynomials of cartesian coordinates without the need to transfom to spherical coordinates. This also means that although the ACE is expanded in spherical harmonics, it takes the form of a polynomial expansion in cartesian coordinates.

As already elucidated in Ref.~\onlinecite{Drautz19}, Eq.(\ref{eq:Ytor}) implies that the moments tensor potentials \cite{Shapeev16} may be written exactly in the form of an ACE. Furthermore, as the ACE provides a basis, while the contraction of the cartesian tensors for the MTPs is to some extent arbitrary, the re-expansion of the MTPs in the form of an ACE may be used to ensure a complete set of basis functions for the moments tensor potentials.

Furthermore, the SOAP descriptor \cite{Bartok2013} is parameterized using hyperspherical harmonics for the Spectral Neighbor Analysis method Potential (SNAP)\cite{Thompson15}. By decomposing the hypershpherical harmonics into a product of an effective radial contribution and a spherical harmonics, one can rewrite the SNAP exactly in the form of an ACE. The details are given in App.~\ref{app:SNAP}.

\section{Multi-component materials}

Before adding magnetic or charge degrees of freedom, I discuss the expansion of scalar, vectorial or tensorial properties in multi-component materials. I assume that the state of a multi-component material is completely characterized by the atomic positions and their chemical species, such that
\begin{equation}
\sigma_j = ( \mu_j, \pmb{r}_{ji}) \,. \label{eq:sigmamulti}
\end{equation}

Next I choose basis functions that are localized on the atoms and are written as a product of chemical, radial and angular contributions
\begin{equation}
\pmb{\phi}_{i \mu_i \kappa n l} (\sigma_j)  =   \pmb{e}_{\kappa}(\mu_j) R^{\mu_j \mu_i}_{nl}(r_{ji}) \pmb{Y}_{l}(\hat{\pmb{r}}_{ji}) \,.
\end{equation}
The basis functions are vectors with elements $\{\pmb{\phi}_{i \mu_i \kappa n l} (\sigma_j)\}_m$, $m = -l,\dots ,l$. Different from Ref.~\onlinecite{Sanchez84}, where the chemical space is expanded in Chebyshev polynomials, I simply use an explicitly orthogonal basis. The $M$ different chemical species are identified by $M$ orthogonal unit vectors in an $M$-dimensional space,
\begin{equation}
\pmb{e}_{\kappa}({\mu}) = \delta_{\kappa \mu} \,,
\end{equation}
such that Eqs.(\ref{eq:orth1},\ref{eq:comp1}) are given by
\begin{align}
\braket{\pmb{e}_{\kappa_1}({\mu})}{ \pmb{e}_{\kappa_2}({\mu})} &= \delta_{\kappa_1 \kappa_2} \,, \\
\sum_{\kappa} \pmb{e}_{\kappa}({\mu_1}) \pmb{e}_{\kappa}({\mu_2}) &= \delta_{\mu_1 \mu_2} \,.
\end{align} 
This has the advantage that chemical species may be added or removed to the system without modifying basis functions of other species and therefore the chemistry dependent expansion coefficients are directly transferable between different materials systems. One may argue that this contradicts the spirit of the original cluster expansion that requires $\phi_0 = 1$. This may easily be taken into account by introducing explicitly a further species, the 'vacuum species' for which $\phi_0 = 1$ and which has no properties associated to it.

The radial functions $R^{\mu_j \mu_i}_{nl}(r_{ji}) $ depend on the distance $r_{ji}$ between the atoms of chemical species $\mu_i$ and $ \mu_j$, while $n$ and $l$ are further indices and $l$ makes reference to the irreducible representation of the rotation group. Evidently, the radial functions are invariant with respect to rotation.

The angular functions  $\pmb{Y}_{l}(\hat{\pmb{r}})$ depend only on the bond direction $\hat{\pmb{r}}$. They form a complete basis for the irreducible representation $l$ of the rotation group, which means that  ${Y}_{l}^{m}(\hat{\pmb{r}})$ with $m = -l, -l+1, \dots, l-1 ,l$ is a vector of $2l+1$ linearly independent basis functions. Typically the angular functions $\pmb{Y}_{l}$ are taken as spherical harmonics, but other, related representations are also possible.

The energy or other configuration dependent quantities are obtained by inserting the basis functions into Eq.(\ref{eq:mbint2}). The atomic base Eq.(\ref{eq:AB}) reads
\begin{equation}
\pmb{A}_{i \mu n l} =  \sum_j  \delta_{\mu \mu_j} R^{\mu_j \mu_i}_{nl}(r_{ji})  \pmb{Y}_{l}(\hat{\pmb{r}}_{ji}) \,,
\end{equation}
which means that in the sum over neigbors $j$ only atoms of species $\mu$ are considered, and 
\begin{equation}
{A}^{(0)}_{i \mu} =  \delta_{\mu \mu_i} \,.
\end{equation}
The expansion of a configuration dependent quantity $\pmb{G}_{i}$ on atom $i$ with species $\mu_i$ that transforms according to the irreducible representation $L_R$ of the rotation group is then written as
\begin{align}
\pmb{G}_{i} = \pmb{G}(\pmb{\sigma}) &= \sum_{\mu n}  c^{(1)}_{\mu_i \mu {n } L_R} \pmb{A}_{i \mu n L_R} \nonumber \\
&+ \sum_{\pmb{\mu n l }}' c^{(2)}_{ \mu_i \pmb{\mu n l} L_R}  
\left(
\begin{array}{c} 
\pmb{l} L_R
\end{array}
\right)_2
\pmb{A}_{i \mu_1 n_1 l_1}    \pmb{A}_{i \mu_2 n_2 l_2}  \nonumber \\
&+ \sum_{\pmb{ \mu n l L}}' 
c^{(3)}_{ \mu_i \pmb{ \mu n l L} L_R}  
 \left(
\begin{array}{c} 
\pmb{l} \\
\pmb{L}
\end{array}
L_R
\right)_3 \nonumber \\
& \phantom{\sum \sum} \times
\pmb{A}_{i \mu_1 n_1 l_1}    \pmb{A}_{i \mu_2 n_2 l_2 }   \pmb{A}_{i \mu_3 n_3 l_3 } \nonumber \\
&+ \sum_{\pmb{ \mu n l L}}' 
c^{(4)}_{ \mu_i \pmb{ \mu n l L} L_R}  
 \left(
\begin{array}{c} 
\pmb{l} \\
\pmb{L}
\end{array}
L_R
\right)_4 \nonumber \\
& \phantom{\sum \sum} \times \pmb{A}_{i \mu_1 n_1 l_1}    \pmb{A}_{i \mu_2 n_2 l_2 }   \pmb{A}_{i \mu_3 n_3 l_3 } \pmb{A}_{i \mu_4 n_4 l_4}  \nonumber \\
&+ \dots \,. 
\end{align}
The sums are taken over lexicographically ordered combinations $\pmb{ \mu n l}$ and the intermediate couplings $\pmb{L}$ which are necessary for a complete set of basis functions. The summation over possible combinations $\pmb{m}$ is implied.

One may define the irreducible set of basis functions of the atomic cluster expansion 
\begin{equation}
\pmb{B}^{(N)}_{\mu_i \pmb{ \mu n l L} L_R} = 
 \left(
\begin{array}{c} 
\pmb{l} \\
\pmb{L}
\end{array}
L_R
\right)_N
\prod_{k = 1}^{N} \pmb{A}_{i \mu_k n_k l_k} \,, \label{eq:ACEB}
\end{equation}
and rewrite the atomic cluster expansion as
\begin{equation}
 \pmb{G}_i= \sum_{N = 0} \sum_{\pmb{ \mu n l L}}'  c^{(N)}_{ \mu_i \pmb{ \mu n l L} L_R} \pmb{B}^{(N)}_{\mu_i \pmb{ \mu n l L} L_R} \,.
\end{equation}
For scalar properties $L_R=0$ the expression may be simplified considerably as this constrains possible intermediate couplings $\pmb{L}$. Further, parity requires $(-1)^{L_R} = (-1)^{\sum_i l_i}$ and selection rules for the couplings apply, Sec.~\ref{sec:trip}.

In complete analogy to the generalization of the SOAP descriptor discussed in Ref.~\onlinecite{Drautz19}, the generalization of ACE to vectorial or tensorial properties may be viewed as providing a generalization of the $\lambda$-SOAP descriptor\cite{Grisafi18} to multi-body interactions.

\section{Including magnetism}

Often other degrees of freedom than chemical species and atomic positions are relevant. I use an atomic cluster expansion that includes magnetism to demonstrate the coupling to other degrees of freedom. In addition to chemical species and position each atom $j$ is assigned a magnetic moment vector $\pmb{m}_j$, such that the configuration of atom $j$ for the expansion on atom $i$ is described by
\begin{equation}
\sigma_j = ( \mu_j, \pmb{r}_{ji}, \pmb{m}_j) \,. \label{eq:sigmaimultim}
\end{equation}
Translational invariance means that only the chemical species and the magnetic moment of atom $i$ enters, but not its position,
\begin{equation}
\sigma_i = ( \mu_i, \pmb{m}_i) \,. \label{eq:sigmajmultim}
\end{equation}

I keep the non-magnetic contributions as they were and expand the magnetic contributions in a radial and angular part,
\begin{equation}
\pmb{\phi}_{i \mu_i \kappa n l n' l'} (\sigma_j)  =   \pmb{e}_{\kappa}(\mu_j) R^{\mu_j \mu_i}_{nl}(r_{ji}) \pmb{Y}_{l}(\hat{\pmb{r}}_{ji}) M^{\mu_j \mu_i}_{n'l'}(m_{j}) \pmb{Y}_{l'}(\hat{\pmb{m}}_{j})  \,,
\end{equation}
where the functions $M$ only depend on the magnitude $m_j$ of the magnetic moment and $\hat{\pmb{m}}_{j}$ is the direction of the magnetic moment. The basis functions are matrices with elements $\{\pmb{\phi}_{i \mu_i \kappa n l n' l'} (\sigma_j)\}_{m m'}$, $m = -l,\dots ,l$ and $m' = -l',\dots ,l'$. 

The atomic cluster expansion is obtained by inserting the basis functions into Eq.(\ref{eq:mbint2}). The atomic base Eq.(\ref{eq:AB}) reads
\begin{equation}
\pmb{A}_{i \mu n l n' l'} =  \sum_j  \delta_{\mu \mu_j} R^{\mu_j \mu_i}_{nl}(r_{ji})  \pmb{Y}_{l}(\hat{\pmb{r}}_{ji}) M^{\mu_j \mu_i}_{n'l'}(m_{j}) \pmb{Y}_{l'}(\hat{\pmb{m}}_{j}) \,,
\end{equation}
and
\begin{equation}
\pmb{A}^{(0)}_{i \mu_i n l} =  M^{\mu_i}_{nl}(m_{i}) \pmb{Y}_{l}(\hat{\pmb{m}}_{i}) \,.
\end{equation}

In the following I choose to separate the angular coupling of atomic positions from the angular coupling of the magnetic moments. As will become clear in the following, this will simplify the expressions without spin-orbit coupling.  The expansion of an atomic quantity that transforms according to the irreducible representation $L_ R$ of the rotation group is then written as
\begin{widetext}
\begin{align}
&\pmb{G}_{i} = \pmb{G}(\pmb{\sigma}) = \sum_{n'}  c^{(0)}_{\mu_i n' L_R} \pmb{A}^{(0)}_{i \mu_i n' L_R}  \nonumber \\
&+\sum_{\mu  {n'_0 n'_1 l'_0 l'_1 L'_I n_1 L_I }}  c^{(1)}_{\mu_i \mu {n'_0 n'_1 l'_0 l'_1 L'_I n_1 L_I } L_R}      (L_I (l'_0 l'_1 L'_I)_2) L_R   \,  \pmb{A}^{(0)}_{i \mu_i n'_0 l'_0} \pmb{A}_{i \mu n_1 L_I n'_1 l'_1} \nonumber \\
&+\sum_{ \pmb{\mu} \pmb{n}  \pmb{l} \pmb{n}'  \pmb{l}' \pmb{L}' L_I L'_I }  c^{(2)}_{\mu_i \pmb{\mu} \pmb{n}  \pmb{l} \pmb{n}'  \pmb{l}' \pmb{L}' L_I L'_I L_R}      
\left((l_1 l_2 L_I)_2 
\left(
\begin{array}{c} 
\pmb{l}' \\
\pmb{L}'
\end{array}
L'_I
\right)_3
\right) 
L_R   \, \pmb{A}^{(0)}_{i \mu_i n'_0 l'_0} \pmb{A}_{i \mu_1 n_1 l_1 n'_1 l'_1} \pmb{A}_{i \mu_2 n_2 l_2 n'_2 l'_2} \nonumber \\
&+\sum_{\pmb{\mu} \pmb{n}  \pmb{l} \pmb{n}'  \pmb{l}' \pmb{L} \pmb{L}' L_I L'_I }  c^{(3)}_{\mu_i \pmb{\mu} \pmb{n}  \pmb{l} \pmb{n}'  \pmb{l}' \pmb{L} \pmb{L}' L_I L'_I  L_R}      
\left(
\left(
\begin{array}{c} 
\pmb{l} \\
\pmb{L}
\end{array}
L_I
\right)_3
\left(
\begin{array}{c} 
\pmb{l}' \\
\pmb{L}'
\end{array}
L'_I
\right)_4
\right) 
L_R   \, \pmb{A}^{(0)}_{i \mu_i n'_0 l'_0} \pmb{A}_{i \mu_1 n_1 l_1 n'_1 l'_1} \pmb{A}_{i \mu_2 n_2 l_2 n'_2 l'_2} \pmb{A}_{i \mu_3 n_3 l_3 n'_3 l'_3} \nonumber \\
&+\sum_{\pmb{\mu} \pmb{n}  \pmb{l} \pmb{n}'  \pmb{l}' \pmb{L} \pmb{L}' L_I L'_I }  c^{(4)}_{\mu_i \pmb{\mu} \pmb{n}  \pmb{l} \pmb{n}'  \pmb{l}' \pmb{L} \pmb{L}' L_I L'_I  L_R}      
\left(
\left(
\begin{array}{c} 
\pmb{l} \\
\pmb{L}
\end{array}
L_I
\right)_4
\left(
\begin{array}{c} 
\pmb{l}' \\
\pmb{L}'
\end{array}
L'_I
\right)_5
\right) 
L_R   \, \pmb{A}^{(0)}_{i \mu_i n'_0 l'_0} \pmb{A}_{i \mu_1 n_1 l_1 n'_1 l'_1} \pmb{A}_{i \mu_2 n_2 l_2 n'_2 l'_2} \pmb{A}_{i \mu_3 n_3 l_3 n'_3 l'_3} \pmb{A}_{i \mu_4 n_4 l_4 n'_4 l'_4} \nonumber \\
&+ \dots \,. \label{eq:generalmagnetic}
\end{align}
\end{widetext}
Here the intermediate resulting angular momenta of the atomic and magnetic system alone are denoted by $L_I$ and $L'_I$, respectively. These two angular momenta couple to the total resulting angular momentum $L_R$. The angular momenta contributions of the atomic part run from $l_1, l_2, \dots$, whereas the magnetic angular moments run from $l'_0, l'_1, l'_2, \dots$. The sums are taken over lexicographically ordered $\pmb{\mu} \pmb{n}  \pmb{l} \pmb{n}'  \pmb{l}'$ (only $n'_0$ and $l'_0$ not ordered) and the intermediate couplings $\pmb{L} \pmb{L}' L_I L'_I $ that are necessary for a complete basis. The summation over possible combinations $\pmb{m}$ and  $\pmb{m}'$ is implied.
 
In analogy to Eq.(\ref{eq:ACEB}) one may define the irreducible set of basis functions of the atomic cluster expansion 
\begin{align}
&\pmb{B}^{(N)}_{\mu_i \pmb{ \mu n l L}  \pmb{n' l' L'} L_I L'_I  L_R} = \nonumber \\
&\left(
\left(
\begin{array}{c} 
\pmb{l} \\
\pmb{L}
\end{array}
L_I
\right)_N
\left(
\begin{array}{c} 
\pmb{l}' \\
\pmb{L}'
\end{array}
L'_I 
\right)_{N+1} \right) L_R \,
\pmb{A}^{(0)}_{i \mu_i n'_0 l'_0} \prod_{k = 1}^{N} \pmb{A}_{i \mu_k n_k l_k n'_k l'_k} \,,
\end{align}
and rewrite the atomic cluster expansion as
\begin{equation}
 \pmb{G}_i = \sum_{N = 0} \sum_{\pmb{ \mu n l L} \pmb{n' l' L'} L_I L'_I }'  c^{(N)}_{ \mu_i \pmb{ \mu n l L} \pmb{n' l' L'} L_I L'_I  L_R} \pmb{B}^{(N)}_{\mu_i \pmb{ \mu n l L} \pmb{n' l' L'} L_I L'_I L_R} \,.
\end{equation}
For the expansion of the energy the resulting angular momentum needs to be zero for rotational invariance, $L_{R}=0$, and therefore $L_I = L'_I$. 

\subsection{Neglecting spin-orbit coupling in the magnetic energy}

The number of coefficients in Eq.(\ref{eq:generalmagnetic}) may appear daunting, but may be structured in a hierarchical way by taking into account that magnetism contributes only a fraction to the cohesive energy. As a very rough estimate one may assume that the contribution of magnetism to the cohesive energy is about 10\%. The change of the magnetic contribution to the energy upon rotation of the magnetic moments is a fraction of the magnetic energy and the contribution of spin-orbit coupling to the magnetic energy is typically orders of magnitude smaller.

If spin-orbit coupling is neglected for the expansion of the energy, then the energy has to be invariant with respect to rotations of the atoms alone at fixed magnetic moments as well as the rotation of the magnetic moments at fixed atomic positions. This means that in Eq.(\ref{eq:generalmagnetic})
\begin{equation}
L_I = L'_I = L_{R} = 0 \,,
\end{equation}
which considerably reduces the number of parameters.
For example, the lowest order interaction term with expansion coefficients $c^{(1)}$ requires $l'_0 = l'_1$ and $l_1 = 0$. A distance dependent Heisenberg interaction is obtained for $l'_0 = l'_1 = 1$, and distance dependent contributions with more complex pairwise angular dependence for  $l'_0 = l'_1 = k$, see also Ref.~\onlinecite{Drautz04-5}.

\subsection{Spin-polarized models and charge transfer}

In spin-polarized models the atomic magnetic moments are confined to point along the $z$-direction. This means that the angular contributions of the magnetic basis functions are limited and one can exploit $\pmb{Y}_{l}(-\hat{\pmb{m}}) = (-1)^{l}\pmb{Y}_{l}(\hat{\pmb{m}})  $, or introduce a discrete basis as in the Ising model or the original cluster expansion~\cite{Sanchez84,Drautz04-5}. Furthermore, magnetic moments that point only along the positive or negative $z$-axis are formally equivalent to positive and negative atomic charges, thus a general atomic cluster expansion that includes charge transfer is also obtained. 

\section{Gradients}

Gradients with respect to the degrees of freedom may be obtained in analogy to Ref.~\onlinecite{Drautz19}. For an efficient numerical implementation it may be advisable to sum over the intermediate couplings first. By combining the expansion coefficients with the generalized Clebsch-Gordan coefficient, summations may be carried out more efficiently,
\begin{widetext}
\begin{align}
&\tilde{c}^{(0)}_{\mu_i n' L_R} = c^{(0)}_{\mu_i n' L_R} \,, \nonumber \\
& \tilde{c}^{(1)}_{\mu_i \mu {n'_0 n'_1 l'_0 l'_1 n_1 } L_R}  =  \sum_{L_I L'_I } c^{(1)}_{\mu_i \mu {n'_0 n'_1 l'_0 l'_1 L'_I n_1 L_I } L_R}      (L_I (l'_0 l'_1 L'_I)_2) L_R   \,,\nonumber \\
&  \tilde{c}^{(2)}_{\mu_i \pmb{\mu} \pmb{n}  \pmb{l} \pmb{n}'  \pmb{l}' L_R}  = \sum_{\pmb{L}' L_I L'_I } c^{(2)}_{\mu_i \pmb{\mu} \pmb{n}  \pmb{l} \pmb{n}'  \pmb{l}' \pmb{L}' L_I L'_I L_R}      
\left((l_1 l_2 L_I)_2 
\left(
\begin{array}{c} 
\pmb{l}' \\
\pmb{L}'
\end{array}
L'_I
\right)_3
\right) 
L_R   \,,\nonumber \\
& \tilde{c}^{(N)}_{\mu_i \pmb{\mu} \pmb{n}  \pmb{l} \pmb{n}'  \pmb{l}'  L_R}    = \sum_{ \pmb{L} \pmb{L}' L_I L'_I } c^{(N)}_{\mu_i \pmb{\mu} \pmb{n}  \pmb{l} \pmb{n}'  \pmb{l}' \pmb{L} \pmb{L}' L_I L'_I  L_R}      
\left(
\left(
\begin{array}{c} 
\pmb{l} \\
\pmb{L}
\end{array}
L_I
\right)_N
\left(
\begin{array}{c} 
\pmb{l}' \\
\pmb{L}'
\end{array}
L'_I
\right)_{N+1}
\right) 
L_R   \,.
 \label{eq:ctilde}
\end{align}
\end{widetext}
The atomic cluster expansion is then written as
\begin{equation}
\pmb{G}_{i} = \sum_{N = 0}  \sum_{\pmb{\mu} \pmb{n}  \pmb{l} \pmb{n}'  \pmb{l}' } \tilde{c}^{(N)}_{\mu_i \pmb{\mu} \pmb{n}  \pmb{l} \pmb{n}'  \pmb{l}'  L_R} \,  \pmb{A}^{(0)}_{i \mu_i n'_0 l'_0} \prod_{k = 1}^{N} \pmb{A}_{i \mu_k n_k l_k n'_k l'_k} \,.
\end{equation}
The derivatives with respect to the atomic base are easily obtained and summarized as
\begin{align}
\pmb{\omega}_{i\mu n l n' l'} &= \pder{\pmb{G}_{i}}{\pmb{A}_{i \mu n l n' l'}} \,, \\
\pmb{\omega}^{(0)}_{i\mu_i n'_0 l'_0} &= \pder{\pmb{G}_{i}}{\pmb{A}^{(0)}_{i \mu_i n'_0 l'_0}} \,, 
\end{align}
and the gradients are expressed as
\begin{align}
\partial_k \sum_i \pmb{G}_{i} &= \sum_i \sum_{\mu n l n' l'} \pmb{\omega}_{i\mu n l n' l'} \partial_k \pmb{A}_{i \mu n l n' l'} \nonumber \\
&+ \sum_{n' l'} \pmb{\omega}^{(0)}_{k\mu_k n' l'} \partial_k \pmb{A}^{(0)}_{k \mu_k n' l'} \,.
\end{align} 
The gradients of the atomic base depend on the degrees of freedom under consideration. In the following I will discuss the gradients with respect to changes to the atomic positions and to the magnetic moments, respectively.

\subsection{Forces}

For the force gradients one has
\begin{align} 
&\nabla_k \pmb{A}_{i \mu n l n' l'} = \sum_j \delta_{\mu \mu_j} \nabla_k \pmb{\phi}_{ \mu_i \mu_j n l n' l'}(\pmb{r}_{ji}) \nonumber \\
&= \delta_{\mu \mu_k} \nabla_k \pmb{\phi}_{ \mu_i \mu_k n l n' l'}(\pmb{r}_{ki}) +  \sum_j \delta_{\mu \mu_j} \nabla_i \pmb{\phi}_{ \mu_i \mu_j n l n' l'}(\pmb{r}_{ji}) \delta_{ik} \nonumber \\
&= \delta_{\mu \mu_k} \nabla_k \pmb{\phi}_{ \mu_i \mu_k n l n' l'}(\pmb{r}_{ki}) -  \sum_j \delta_{\mu \mu_j} \nabla_j \pmb{\phi}_{ \mu_j \mu_i n l n' l'}(\pmb{r}_{ji}) \delta_{ik} \,,
\end{align}
where I made use of
\begin{align}
\pmb{\phi}_{ \mu_i \mu_j n l n' l'}(\pmb{r}_{ji}) &= \pmb{\phi}_{ \mu_j \mu_i n l n' l'}(\pmb{r}_{ji}) \,, \\
\nabla_i \pmb{\phi}_{ \mu_i \mu_j n l n' l'}(\pmb{r}_{ji}) &= -\nabla_j \pmb{\phi}_{ \mu_j \mu_i n l n' l'}(\pmb{r}_{ji}) \,,
\end{align}
and
\begin{equation}
 \nabla_k \pmb{A}^{(0)}_{i \mu_i n' l'} = 0 \,. 
\end{equation}
Then by defining
\begin{equation}
\pmb{f}_{ki} = \sum_{n l n' l'} \pmb{\omega}_{i\mu_k n l n' l'}  \nabla_k \pmb{\phi}_{ \mu_i \mu_k l n' l'}(\pmb{r}_{ki}) \,,
\end{equation}
the gradient is written as
\begin{equation}
\pmb{F}_k = \sum_i - \nabla_k \pmb{G}_i = \sum_i \left( \pmb{f}_{ik} - \pmb{f}_{ki} \right) \,.
\end{equation}
As in a typical molecular dynamics implementation one evaluates the forces on all atoms and to this end loops over all neighbors for each atom, for the bond $k-i$ only the contribution $\pmb{f}_{ki} $ needs to be evaluated and the contribution $\pmb{f}_{ik}$ can be added when the bond $i-k$ is visited. The pairwise representation of the forces further enables the evalulation of the virial stresses and pressure at no additional computational cost. \cite{Thompson09}

Non-linear functions $F$ of the atomic cluster expansion, {\it i.e.}, $\sum_i F( \pmb{G}_i )$ may be evaluated along the same lines from simple embedding functions to non-linear machine learning representations. \cite{Drautz19} 

\subsection{Magnetic torques}

For the magnetic gradients one has
\begin{equation} 
\pder{}{\pmb{m}_k} \pmb{A}_{i \mu n l n' l'} =\delta_{\mu \mu_k} \pder{}{\pmb{m}_k} \pmb{\phi}_{ \mu_i \mu_k n l n' l'}(\pmb{m}_{k})  \,,
\end{equation}
and
\begin{equation}
\pder{}{\pmb{m}_k} \pmb{A}^{(0)}_{i \mu_i n' l'} =   \delta_{i k}\delta_{\mu_i \mu_k} \pder{}{\pmb{m}_k} \pmb{A}^{(0)}_{k \mu_k n' l'}\delta_{ik} \,. 
\end{equation}
The gradient is therefore written as
\begin{align}
\pder{}{\pmb{m}_k} \sum_i \pmb{G}_i &= \sum_i \sum_{ n l n' l'} \pmb{\omega}_{i\mu_k n l n' l'}  \, \pder{}{\pmb{m}_k} \pmb{\phi}_{ \mu_i \mu_k n l n' l'}(\pmb{m}_{k})  \nonumber \\ 
&+ \sum_{ n' l'} \pmb{\omega}^{(0)}_{k\mu_k n' l'} \pder{}{\pmb{m}_k} \pmb{A}^{(0)}_{k \mu_k n' l'} \,.
\end{align}
The gradient comprises contributions due to transversal as well as longitudinal changes of the atomic magnetic moments. Most spin-dynamics implementations assume that the magnitude of the spins is preserved during simulations and including longitudinal degrees of freedom requires a modified equation of motion \cite{Ma08,Tranchida18,Ma12}.

\section{Conclusions}

Quantitative predictions for the development and design of novel materials require models of the interatomic interaction that may be converged systematically to represent reference data with arbitrary precision and that at the same time are numerically efficient for sampling or large scale atomistic simulations. The atomic cluster expansion provides efficient expressions for the evaluation of atomic scale properties. Here I extended the atomic cluster expansion to vectorial and tensorial properties and to include degrees of freedom such as atomic magnetic moments and charges in addition to the atomic positions.

The resulting expressions are a coherent extension of the original ACE, with a similar structure and therefore their implementation does not require significantly more effort. At lowest order contact with simple models may be made, for example, a distance-dependent Heisenberg interaction is obtained for an expansion that takes into account atomic positions and atomic magnetic moments. For a full expansion the number of parameters that need to be fitted to reference data increased significantly, however, clear hierarchies in the interatomic interaction should help to define the parameters in a robust way. In order to fully assess the efficiency of the magnetic ACE, the next step must be its parameterization for a particular system.

\begin{acknowledgments}
I acknowledge helpful discussions with Marc Cawkwell, G\'abor Cs\'anyi, Genevi\`eve Dusson, Yury Lysogorskiy, Christoph Ortner, Matteo Rinaldi and Aidan Thompson and funding through the German Science Foundation (DFG), project number 405621217.
\end{acknowledgments}

\appendix

\section{Expansion of tensor products in spherical harmonics \label{app:MT}}

The Clebsch-Gordan coefficients are unitary. For spherical harmonics this means
\begin{equation}
Y_{L}^M = \sum_{m_1 m_2} C_{L l_1 l_2}^{M m_1 m_2} Y_{l_1}^{m_1}  Y_{l_2}^{m_2} \,, 
\end{equation}
and
\begin{equation}
Y_{l_1}^{ m_1}  Y_{l_2}^{ m_2}  = \sum_{LM} \tilde{C}_{l_1 l_2 L}^{m_1 m_2 M} Y_{L}^{M}\,,
\end{equation}
where only matrix elements with $M = m_1 + m_2$ are different from zero and
\begin{equation}
\tilde{C}_{l_1 l_2 L}^{m_1 m_2 M}  = \sqrt{\frac{(2 l_1 +1)(2l_2 + 1)}{ 4 \pi ( 2 L + 1)}} C_{L l_1 l_2}^{0 0 0} C_{L l_1 l_2}^{M m_1 m_2} \,.
\end{equation}

A unit vector $\hat{\pmb{r}}$ may be expressed as a linear combination of spherical harmonics with $l=1$,
\begin{equation}
\hat{r}_n = \sum_{m = -1}^1 a_{nm} Y_{1}^{m} \,,
\end{equation}
with the transformation matrix
\begin{equation}
{a}_{nm} = \sqrt{ \frac{2\pi}{3} }\left(
\begin{array}{ccc}
-1 & 0 & 1 \\
i & 0 & i \\
0 & \sqrt{2} & 0 
\end{array}
\right)_{nm} \,.
\end{equation}
For example, the matrix elements of $\hat{\pmb{r}} \otimes \hat{\pmb{r}}$ are given by
\begin{align}
&\hat{r}_{n_1} \hat{r}_{n_2} = \sum_{m_1, m_2  = -1}^1 a_{n_1 m_1} a_{n_2 m_2}  Y_{1}^{m_1 } Y_{1}^{m_2 } \nonumber \\
&= \sum_{m_1, m_2  = -1}^1 a_{n_1 m_1} a_{n_2 m_2} \sum_{L = 0,1,2} \sum_{M = -L}^L \tilde{C}_{11L}^{m_1 m_2 M} Y_{L}^{M}  \,.
\end{align}
The matrix elements of $\hat{\pmb{r}} \otimes \hat{\pmb{r}} \otimes \hat{\pmb{r}}$ are given by
\begin{align}
&\hat{r}_{n_1} \hat{r}_{n_2} \hat{r}_{n_3} = \sum_{m_1 m_2 m_3  = -1}^1 a_{n_1 m_1} a_{n_2 m_2}  a_{n_3 m_3}  \times \nonumber \\
 &\sum_{L_1 = 0}^2 \sum_{L_2 = | L_1 - 1 |}^{L_1 + 1}   \sum_{M_1 = -L_1}^{L_1}  \sum_{M_2 = -L_2}^{L_2} \tilde{C}_{11L_1}^{m_1 m_2 M_1} \tilde{C}_{1 L_1 L_2}^{m_3 M_1 M_2} Y_{L_2}^{M_2 } \,.
\end{align}
This is easily generalized to arbitrary order by introducing the transformation matrix
\begin{align}
&X_{n_1 n_2 n_3 \dots n_N }^{L M}= \sum_{m_1 m_2 m_3  \dots m_N = -1}^1 \left( \prod_{k = 1}^N  a_{n_k m_k} \right) \times \nonumber \\
&\sum_{\substack{L_1 L_2 \dots L_{N-2}\\ M_1 M_2 \dots M_{N-2}}} \tilde{C}_{11L_1}^{m_1 m_2 M_1} \left( \prod_{k = 1}^{N-3}  \tilde{C}_{1 L_k L_{k+1}}^{m_{k+2} M_k M_{k+1}} \right) \tilde{C}_{1 L_{N-2} L}^{m_{N} M_{N-2} M }  \,. \label{eq:X}
\end{align}
The transformation matrix is different from zero only for $0 \leq L\leq N$ and $-L \leq M \leq L$ and the product tensors may be written in spherical harmonics as
\begin{align}
\hat{r}_{n_1} \hat{r}_{n_2} \dots \hat{r}_{n_N} &= \sum_{L = 0}^N \sum_{M = -L}^L X_{n_1 n_2 n_3 \dots n_N}^{L M} Y_{L}^{M} \,. 
\end{align}

\section{Spherical harmonics represented as polynomials of cartesian coordinates \label{app:SH} }

\subsubsection{Traditional evaluation of spherical harmonics}

A unit vector $\hat{\pmb{r}}$ of length one is given in spherical coordinates as
\begin{align}
\hat{r}_x &= \sin \theta \cos \phi \,, \\
\hat{r}_y &= \sin \theta \sin \phi \,, \\
\hat{r}_z &= \cos \theta \,.
\end{align}
Traditionally the spherical harmonics are obtained as functions of $\cos \theta$, $\cos \phi$ and  $\sin \phi$, but $\sin \theta$ is also required explicitly for setting up the associated Legendre polynomials or to obtain  $\cos \phi$ and  $\sin \phi$, see Ref.~\onlinecite{Limpanuparb14} for an efficient algorithm. The computation of $\sin \theta$ requires an explicit square root function evaluation,
\begin{equation}
\sin \theta = \sqrt{ 1 - \cos^2 \theta} \,,
\end{equation}
where $\sin \theta >0$ is sufficient. In addition to computational cost for the square root evaluation, a direct naive implementation of the conversion between spherical and cartesian coordinates is numerically unstable when $\sin \theta \approx 0$ and generates a number of potentially redundant floating point operations.

\subsubsection{Evaluation in cartesian coordinates}

As the spherical harmonics fulfill
\begin{equation}
(Y^{m}_{l})^* = (-1)^m Y^{-m}_{l}  \,,
\end{equation}
in the following I consider only $m \geq 0$. The spherical harmonics may be represented as 
\begin{align}
&Y^{m}_{l} = \sqrt{ \frac{ (2l +1) }{4 \pi}  \frac{(l-m)! }{(l+m)!} } \, P_{l}^m(\cos \theta) e^{im \phi} \label{eq:SH}\\ 
&= \sqrt{ \frac{ (2l +1) }{4 \pi}  \frac{(l-m)! }{(l+m)!} } \, (-1)^m e^{im \phi}   (\sin \theta)^m \der{^m}{(\cos \theta)^m} P_{l}(\cos \theta) \,,
\end{align}
with the Legendre polynomials $P_{l}(\cos \theta)$. I define 
\begin{equation}
\bar{P}_{l}^m(\cos \theta) =  (-1)^m \sqrt{ \frac{ (2l +1) }{4 \pi}  \frac{(l-m)! }{(l+m)!} }\der{^m}{(\cos \theta)^m} P_{l}(\cos \theta) \,.
\end{equation}
One then immediately has
\begin{equation}
\hat{r}_x + i\hat{r}_y = e^{i \phi}  \sin \theta \,,
\end{equation}
and the spherical harmonics in explicitly polynomial form and in cartesian coordinates are given by
\begin{equation}
Y^{m}_{l} = (\hat{r}_x + i\hat{r}_y)^m  \bar{P}_{l}^m(\hat{r}_z)  \,.
\end{equation}
Iterative expressions may be obtained trivially as
\begin{equation}
(\hat{r}_x + i\hat{r}_y)^m =  (\hat{r}_x + i\hat{r}_y)^{m-1} \, (\hat{r}_x + i\hat{r}_y) \,,
\end{equation}
and from modifying the expressions used in Ref.~\onlinecite{Limpanuparb14},
\begin{align}
\bar{P}_{l}^l &= c_l \bar{P}_{l-1}^{l-1} \,, \\
\bar{P}^{l}_{l+1} &= d_l \hat{r}_z \bar{P}_{l}^{l} \,, \\
\bar{P}^{m}_{l} &= a_l^m (\hat{r}_z   \bar{P}^{m}_{l-1} + b_l^m \bar{P}^{m}_{l-2} )\,,
\end{align}
with
\begin{align}
\bar{P}_{0}^0 &= \sqrt{\frac{1}{4\pi}} \,,\\
 a_l^m &= \sqrt{\frac{4 l^2 -1}{l^2-m^2}} \,,\\
 b_l^m &= -\sqrt{\frac{(l-1)^2 - m^2}{4(l-1)^2-1}} \,,\\
 c_l &= -\sqrt{1 + \frac{1}{2l}} \,, \\
d_l &= \sqrt{2l + 3} \,.
\end{align}

\subsubsection{Derivatives in cartesian coordinates}

A slight modification of the recursion formulae leads to expressions for the derivatives $d\bar{P}^m_l = \der{\bar{P}^m_l}{ \hat{r}_z}$,
\begin{align}
d\bar{P}_{l}^l &= 0 \,, \\
d\bar{P}^{l}_{l+1} &= d_l \bar{P}_{l}^{l} \,, \\
 d\bar{P}^{m}_{l} &= a_l^m (\bar{P}^{m}_{l-1} + \hat{r}_z   d\bar{P}^{m}_{l-1} + b_l^m d\bar{P}^{m}_{l-2} )\,.
\end{align}
The derivatives of the spherical harmonics are written as
\begin{align}
\pder{Y^m_l}{\hat{r}_x} &= m (\hat{r}_x + i\hat{r}_y)^{m-1}  \bar{P}_{l}^m(\hat{r}_z) \,, \\
\pder{Y^m_l}{\hat{r}_y} &= im (\hat{r}_x + i\hat{r}_y)^{m-1}  \bar{P}_{l}^m(\hat{r}_z) = i \pder{Y^m_l}{\hat{r}_x} \,, \\
\pder{Y^m_l}{\hat{r}_z} &= (\hat{r}_x + i\hat{r}_y)^{m}  \der{\bar{P}_{l}^m(\hat{r}_z)}{\hat{r}_z}  \,. \\
\end{align}
Noting the derivative of the unit length vector
\begin{equation}
\partial_j \hat{r}_i = \delta_{ji} - \hat{r}_j \hat{r}_i \,,
\end{equation}
and combining
\begin{equation}
y^m_l = \hat{r}_x \pder{Y^m_l}{\hat{r}_x} + \hat{r}_y \pder{Y^m_l}{\hat{r}_y} + \hat{r}_z \pder{Y^m_l}{\hat{r}_z} \,,
\end{equation}
one arrives at
\begin{equation}
\partial_i {Y^m_l} = \pder{Y^m_l}{\hat{r}_i} - y^m_l \hat{r}_i \,.
\end{equation}

\section{Spectral neighbor analysis potential expressed as an atomic cluster expansion \label{app:SNAP}}

\subsubsection{Hyperspherical harmonics}

The hyperspherical harmonics may be written as \cite{Mason09}
\begin{align}
&Z^n_{lm}(\omega, \theta, \varphi) = \nonumber \\
& (-i)^l \frac{2^{l+1/2} l! }{2 \pi } \left[ (2l+1) \frac{ (l-m)!}{(l+m)!} \frac{(n+1)(n-l)!}{(n+l+1)!} \right]^{1/2} \nonumber \\
& \times[\sin(\omega/2)]^l C_{n-l}^{l+1} (\cos(\omega/2)) P_l^m( \cos \theta) \exp(i m \varphi) \,,
\end{align}
with integer indices $0\leq n$, $0\leq l \leq n$ and $-l \leq m \leq l$, and
where $C_{n-l}^{l+1}$  is a Gegenbauer polynomial and $P_l^m$ an associated Legendre polynomial. 

A unit vector $\hat{s}$ of length one is given in spherical coordinates as
\begin{align}
\hat{s}_0 &= \cos \omega  \,, \\
\hat{s}_1 &= \sin \omega  \cos \theta \,, \\
\hat{s}_2 &= \sin \omega  \sin \theta \cos \varphi \,, \\
\hat{s}_3 &= \sin \omega  \sin \theta \sin \varphi \,.
\end{align}

\subsubsection{4-dimensional basis of the SOAP descriptor}

In a variant of the SOAP descriptor \cite{Bartok2013} a 3-dimensional vector $\pmb{r} = (x,y,z)$ of length $r$ is mapped onto a 4-dimensional unit sphere by using the transformation
\begin{align}
\varphi      &= \arctan( x/y) \,, \nonumber \\
\theta   &= \arccos( z/r) \,, \nonumber \\
\omega &= \pi r/r_0 \,,
\end{align}
where $r_0$ is larger or equal to the cutoff distance, {\it i.e.}, in a simulation one will have $ r/r_0 \leq 1$.

\subsubsection{Representation in spherical harmonics}

I define particular radial functions as
\begin{align}
R_{nl}(r) &=  (-i)^l 2^{l+1/2} l!\sqrt{\frac{(n+1)(n-l)!}{\pi (n+l+1)!} } \nonumber \\
& \times \left[\sin \left(\frac{\pi}{2} \frac{r}{r_0} \right) \right]^l C_{n-l}^{l+1} \left(\cos \left(\frac{\pi}{2} \frac{r}{r_0}\right) \right) \,. \label{eq:radial}
\end{align}
By just using the basic definition of the spherical harmonics Eq.(\ref{eq:SH}) the hyperspherical harmonics may be represented as
\begin{equation}
Z^n_{lm}(r, \theta, \varphi) =  R_{nl}(r)  Y^{m}_{l} (\theta,\varphi) \,. \label{eq:basis}
\end{equation}
One sees that the hyperspherical harmonics may be viewed as a particular choice of radial basis functions for the ACE
\begin{equation}
\phi_{nlm}(\pmb{r}) = Z^n_{lm}(r, \theta, \varphi) \,,
\end{equation}
with the radial functions $R_{nl}(r)$ given by Eq.(\ref{eq:radial}). 

This means that the SOAP descriptor and therefore the SNAP\cite{Thompson15} can immediately and exactly be rewritten in the from of an ACE. The expansion coefficients for the representation of SNAP in the form of an ACE may be obtained by inserting the expression Eq.(\ref{eq:basis}) into SNAP and reading off the expansion coefficients from the ACE product basis functions.

\end{document}